\documentstyle[12pt,psfig]{article}
\def\simgt{\stackrel{>}{{}_\sim}}
\def\be{\begin{equation}}
\def\ee{\end{equation}}
\def\bear{\be\begin{array}}
\def\eear{\end{array}\ee}
\def\bea{\begin{eqnarray}}
\def\eea{\end{eqnarray}}

\def\baselinestretch{1}
\newcommand{\msn}{m_{ \tilde{\nu} }  }
\newcommand{\msl}{m_{ \tilde{l}_L }  }
\newcommand{\msr}{m_{ \tilde{l}_R }  }

\begin{document}
%%%%%%%%%%%%%%%%%%%%%%%%%%% subequations.sty %%%%%%%%%%%%%%%%%%%%%%%%
\catcode`@=11
\newtoks\@stequation
\def\subequations{\refstepcounter{equation}%
\edef\@savedequation{\the\c@equation}%
  \@stequation=\expandafter{\theequation}%   %only want \theequation
  \edef\@savedtheequation{\the\@stequation}% % expanded once
  \edef\oldtheequation{\theequation}%
  \setcounter{equation}{0}%
  \def\theequation{\oldtheequation\alph{equation}}}
\def\endsubequations{\setcounter{equation}{\@savedequation}%
  \@stequation=\expandafter{\@savedtheequation}%
  \edef\theequation{\the\@stequation}\global\@ignoretrue

\noindent}
\catcode`@=12
%%%%%%%%%%%%%%%%%%%%%%%%%%%%%%%%%%%%%%%%%%%%%%%%%%%%%%%%%%%%%%%%%%%%%
\begin{titlepage}
\title{{\bf
Constraints on supersymmetric theories from $\mu\rightarrow e,\gamma$
}
\thanks{Work partly supported
by CICYT under contract AEN94-0928, and  by the European Union under contract
No. CHRX-CT92-0004.}
}
\author{ {\bf B. de Carlos${}^{ {\#}}$ }\thanks{Supported
by a Spanish MEC Postdoctoral Fellowship.}
,
{\bf J.A. Casas${}^{ {\ddag}}$ }
and {\bf J.M. Moreno${}^{ {\ddag}}$}\\
\hspace{3cm}\\
${}^{\#}$ {\small Dept. of Theoretical Physics} \\
{\small Univ. of Oxford, 1 Keble Road, Oxford OX1 3NP, U.K.}  \\
{\small e--mail: decarlos@thphys.ox.ac.uk} \\
\vspace{-0.3cm}\\
${}^{\ddag}$ {\small Instituto de Estructura de la Materia,
CSIC}\\
{\small Serrano 123, 28006 Madrid, Spain}\\
{\small e--mail: casas@cc.csic.es, imtje30@cc.csic.es} }
\date{}
\maketitle
\def\baselinestretch{1.15}
\begin{abstract}
\noindent
In the absence of any additional assumption it is natural to conjecture
that sizeable flavour-mixing mass entries, $\Delta m^2$, may appear in the
mass matrices of the scalars of the MSSM, i.e. $\Delta m^2\sim O(m^2)$.
This flavour violation can still be reconciled with the experiment
if the gaugino mass, $M_{1/2}$, is large enough, leading
to a {\em gaugino dominance} framework (i.e. $M_{1/2}^2\gg m^2$),
which permits a remarkably model--independent analysis. We study this
possibility focussing our attention on the $\mu\rightarrow e,\gamma$ decay.
In this way we obtain very strong and general constraints, in particular
$\frac{M_{1/2}^2}{\Delta m}
\simgt 34\ {\rm TeV}$.
On the other hand, we show that our analysis and results
remain valid for values of $m^2$ much larger than $\Delta m^2$,
namely for $\frac{\Delta m^2}{m^2}\simgt \frac{m^2}
{10\ {\rm TeV^2}}$, thus extending
enormously their scope of application.
Finally, we discuss the implications for superstring scenarios.

\end{abstract}

\thispagestyle{empty}

\vspace{2cm}
\leftline{}
\leftline{IEM--FT--108/95}
\leftline{OUTP--95--29P}
\leftline{July 1995}
\leftline{}

\vskip-23.3cm
\rightline{}
\rightline{{\bf IEM--FT--108/95}}
\rightline{{\bf OUTP--95--29P}}
\vskip3in

\end{titlepage}
\newpage
%%%%%%%%%%%%%%%%%%%%%%%%%%%%%%%%%%%%%%%%%%%%%%%%%%%%%%%%%%%%%%%%%%%
\setcounter{page}{1}

\section{Introduction}

It is well-known that FCNC processes are very sensitive tests to
physics beyond the standard model (SM) and, in particular, to
supersymmetric extensions of the SM (SSM). There are two main sources
of FCNC signals in the SSM. First, since FCNC processes occur
only beyond tree-level, they are sensitive to the existence of
new particles circulating in the relevant loops. In this way,
the usual SM predictions on FCNC processes (e.g. $K\bar K$
mixing or $b\rightarrow s,\gamma$) are modified by any
SSM \cite{ellis82}. Second, besides the Kobayashi--Maskawa mechanism,
supersymmetry provides new direct sources of flavour violation,
namely the  possible (and even natural as we will see) presence of
{\em off-diagonal} terms (say generically $\Delta m^2$) in the squark
and slepton mass matrices. In the present paper we will
concentrate on this second source of flavour violation, which induces
not only modifications in the SM FCNC processes (such as
$b\rightarrow s,\gamma$), but also the appearance of new FCNC
processes, particularly in the leptonic sector, e.g.
$\mu\rightarrow e,\gamma$ or $\tau\rightarrow e,\gamma$.

In this paper we will focus all our attention on the constraints
on $\Delta m^2$ from the $\mu\rightarrow e,\gamma$ process for
the following reasons:

\begin{description}

\item[{\em i)}]

Among all the FCNC processes (both in the hadronic and
in the leptonic sector) $\mu\rightarrow e,\gamma$ is by far the one with
higher potential to restrict the value of the off-diagonal terms,
$\Delta m^2$ \cite{gabbi89,hagel94,choud95,brax94,barbi94}.

\item[{\em ii)}]

Unlike in the hadronic processes, the calculation
of $\mu\rightarrow e,\gamma$ is very clean and not affected by big
uncertainties.

\end{description}

\noindent
Actually, we will see that the constraints on the
SSM coming from $\mu\rightarrow e,\gamma$ are not only very strong, but
also that its evaluation is remarkably independent of the particular details
of the model under consideration.

Let us briefly review the emergence of non-diagonal scalar mass matrices
in the minimal supersymmetric standard model (MSSM).

The MSSM is defined by the superpotential, $W$, and the form of the
soft supersymmetry breaking terms. $W$ is given by
\be
\label{W}
W=\sum_{i,j}\left\{
h^u_{ij}Q_i H_2 {u_R}_j + h^d_{ij}Q_i H_1 {d_R}_j
+ h^e_{ij}L_i H_1 {e_R}_j \right\} +  \mu H_1 H_2\;\; ,
\ee
where $i,j$ are generation indices, $Q_i$ ($L_i$) are the scalar
partners of the quark (lepton) SU(2) doublets,
${u_R}_i,{d_R}_i$ (${e_R}_i$) are the quark (lepton) singlets and $H_{1,2}$
are the two SUSY Higgs doublets; the $h_{ij}$--factors are the
(matricial )Yukawa
couplings and $\mu$ is the usual Higgs mixing parameter. In all terms
of eq.~(\ref{W}) the usual SU(2) contraction is assumed,
e.g. $H_1H_2\equiv\epsilon_{\alpha\beta} H_1^\alpha H_2^\beta$ with
$\epsilon_{12}=-\epsilon_{21}=1$.
{}From $W$ the
(global) supersymmetric part of the Lagrangian, ${\cal L}_{\rm SUSY}$, is
readily obtained.
\bea
\label{Lsusy}
{\cal L}_{\rm SUSY}&=&-\sum_k\left|\frac{\partial W}{\partial
\phi_k}\right|^2
-\frac{1}{2}\left(\sum_{k,l}\left[\frac{\partial^2 W}{\partial
\phi_k \partial \phi_l} \right] \psi_k\psi_{l}
 + {\rm h.c.}\right)\ +\ {\rm D-terms}
\nonumber \\
&+& {\rm gauge}\;\ {\rm interactions}\;\;,
\eea
where $\phi_{k,l}$ ($\psi_{k,l}$) run over all the scalar (fermionic)
components of the chiral superfields.

In addition to this, the soft breaking terms
(gaugino and scalar masses, and trilinear and bilinear scalar terms) coming
from the (unknown) supersymmetry breaking mechanism have the form
\bea
-{\cal L}_{\rm soft}&=&\frac{1}{2}M_a\lambda_a\lambda_a +
\left(m_L^2\right)_{ij} \bar L_iL_j\ +\
\left(m_{e_R}^2\right)_{ij} \bar{e_R}_i {e_R}_j\
\nonumber \\
&+&\
\left(m_Q^2\right)_{ij} \bar Q_iQ_j\ +\
\left(m_{u_R}^2\right)_{ij} \bar {u_R}_i {u_R}_j\ +\
\left(m_{d_R}^2\right)_{ij} \bar {d_R}_i {d_R}_j
\label{nodiag} \\
&+&\  \left[A^u_{ij}h^u_{ij}Q_i H_2 {u_R}_j +
A^d_{ij}h^d_{ij}Q_i H_1 {d_R}_j
+ A^e_{ij}h^e_{ij}L_i H_1 {e_R}_j + B \mu H_1H_2 +{\rm h.c.} \right]\ ,
\nonumber
\eea
where $a$ is a gauge group index, $\lambda_a$ are the gauginos,
and the remaining fields in the formula denote just their corresponding
scalar components. In the simplest version of the MSSM the
soft breaking parameters are taken as {\em universal} (at the unification
scale $M_X$). Then, the independent parameters of the theory are
\be
\label{inPar}
\mu,m,M_{1/2},A,B
\ee
(the rest of the parameters can be worked out demanding a correct
unification of the gauge coupling constants and correct masses for all
the observed particles). However this simplification is not at all a
general principle. In particular there is no theoretical argument
against $m^2_{ij}$, $A_{ij}$ having a non-diagonal structure (as
reflected in eq.~(\ref{nodiag})). The
ultimate reason for the specific pattern of these matrices
has to be searched in the type of SUGRA theory from which the MSSM
derives (particularly in the structure of the K\"ahler potential) and on
the mechanism of supersymmetry breaking. Both ingredients are
nowadays unknown, although superstring theories offer plausible
ansatzs, which, by the way, do not support in general the universality
assumption
(we will turn to this point later). On the other hand, it is expected
that the Yukawa matrices, $h^u_{ij}$, $h^d_{ij}$, $h^e_{ij}$, may
be also non-diagonal (this is actually obliged for $h^u_{ij}$ or
$h^d_{ij}$). After performing the usual rotation in the
superfields to diagonalize\footnote{We will work from now on in this
new basis for the superfields.} $h^u_{ij}$, $h^d_{ij}$, $h^e_{ij}$,
the fermionic mass matrices and the contribution to the scalar mass
matrices coming from\footnote{This contribution only takes place once
$\langle H_1\rangle$, $\langle H_2\rangle$ $\neq 0$ and, except for the
stop mass, is much smaller than the mass terms of eq.~(\ref{nodiag}).}
(\ref{Lsusy}) become diagonal. However,
if the $m^2_{ij}$ matrices are
not universal, they will develop off-diagonal entries, even if they
are initially diagonal. A similar thing occurs with the $A_{ij}$ matrices.
In the following we will concentrate our attention on the
$(m^2_{ij})_{i\neq j}\equiv \Delta m^2_{ij}$ terms,
setting $(A_{ij})_{i\neq j}=0$. Notice here that $A_{ij}$ always comes
accompanied by the corresponding Yukawa coupling (see eq.~(\ref{nodiag}) or
eqs.~(\ref{rgesc},\ref{rgestr}) below). This makes its effect negligible
when studying the impact of $\Delta m^2_{ij}$ on $\mu\rightarrow e,\gamma$.
On the other, hand $(A_{ij})_{i\neq j}$ is also constrained by FCNC processes
\cite{gabbi89,choud95}
(and in particular by $\mu\rightarrow e,\gamma$ itself), but we will not
be concerned by this issue in the present work.

{}From the previous arguments,
it is natural to assume that the off-diagonal entries, $\Delta m^2$, can
be {\em sizeable}, or even of the same order as the diagonal terms
\be
\label{sizedelt}
\Delta m^2 \sim O(m^2)\;\;.
\ee
Certainly, there are proposed mechanisms to avoid this, e.g. the
above-mentioned assumption of universality, horizontal symmetries
\cite{nir93,dine93},
plastic soft terms \cite{dimop95p}, etc. On the other hand, in some
string scenarios \cite{decar93p,kaplu93,brign94}
$\Delta m^2$ could naturally be very small. However one should wonder
whether, in the absence of any additional assumption, the perfectly possible
and even natural situation of eq.~(\ref{sizedelt}) could still be compatible
with the experimental data and, more precisely, with the present experimental
bound \cite{bolto88} on $\mu\rightarrow e,\gamma$
\be
\label{megexp}
BR(\mu\rightarrow e,\gamma)\le 5\times10^{-11}\;\;,
\ee
which, as stated above, gives the strongest constraints on $\Delta m^2$,
more precisely on $\Delta m^2_{e\mu}$

This will be the main goal of this article. The results will turn out
to be surprisingly model-independent. In fact they will depend only on
two parameters: the value of $\Delta m^2$ itself and the gaugino mass,
$M_{1/2}$. Also, we will make more precise (and relaxed) the assumption of
eq.~(\ref{sizedelt}), namely our results will be valid for any $\Delta m^2$
satisfying
\be
\label{sizedelt2}
\frac{\Delta m^2}{m^2}\simgt \frac{m^2}{10\ {\rm TeV^2}}\;.
%\frac{\Delta m}{m} \simgt \frac{m}{3\ {\rm TeV}} \;\;.
\ee
In principle it might seem that
$\Delta m^2_{e\mu}=O(m^2_{e,\mu})$
is absolutely incompatible with (\ref{megexp}) due to the existence of
diagrams of the type of those shown in Fig.~1 (to be discussed in detail
in sect.~3). However this is not necessarily the case, since
$\mu\rightarrow e,\gamma$ is a low energy process, while the assumption
(\ref{sizedelt}) naturally arises at some unification scale,
$\sim M_X$.
%where the supersymmetry breaking effects take place.
Hence, an ordinary renormalization group (RG) running is in order.
As a matter of fact, this running softens the problem, since the
diagonal entries of the mass matrices increase
substantially \cite{dine90,brign94} while the off--diagonal ones
remain almost unchanged.
This possibility has also been considered in ref.~\cite{choud95},
though the calculation was affected by very large uncertainties,
as we will see afterwards.

In sect.~2 the RG running from high to low energies is carried out,
stressing the fact that the $\Delta m^2\sim O(m^2)$ conjecture,
eq.~(\ref{sizedelt}), automatically leads
to a ``gaugino dominance" framework (i.e. $M_{1/2}^2\gg m^2$),
where all the relevant low-energy quantities
apart from $\Delta m^2$ itself are determined in terms of a
unique parameter, $M_{1/2}$. In sect.~3 the complete calculation
of $BR(\mu\rightarrow e,\gamma)$ is performed, indicating
which are the most important diagrams and why. The results are presented in
sect.~4. They are very general, depending only on the values of
$M_{1/2}$ and
$\Delta m$, and have strong implications, in particular the need
of a large and precise $M_{1/2}^2/\Delta m^2$ hierarchy
in order to maintain the consistency with the experiment.
Furthermore, it will be shown that the analysis and results are
valid for a wide range of $\Delta m^2/m^2$ values,
as anticipated in
eq.~(\ref{sizedelt2}). Finally, the summary and conclusions are presented
in sect.~5.

\section{From high to low energies}

The expression of $BR(\mu\rightarrow e,\gamma)$ in the MSSM depends
on several low-energy parameters, namely $\tan\beta\equiv\langle
H_2\rangle/\langle H_1\rangle$, $\mu$, $A$, and the
spectrum of masses of sleptons and gauginos (the details of the precise
dependence are left for sect.~3). These are not independent parameters.
They come as a low energy result of the actual initial parameters
of the theory (see eq.~(\ref{inPar})) , which are assumed to be
given at the unification scale, $M_X$. Of course, the connection
has to be made through the corresponding RGEs. Next we list some
of the most relevant RGEs in our calculation \cite{deren84}:

\begin{itemize}

\item Gaugino masses:
%Gauginos
\be
\frac{d M_a}{dt}  =  -b_a \tilde{\alpha}_a M_a \; , \;\;\;\;  a= 1,2,3
\label{rgegau}
\ee
where $t = 2 \log(M_X/Q)$, $Q$ being the scale at which we evaluate the
masses, $b_{a}$ are the coefficients of the 1--loop beta functions for the
gauge couplings, $\tilde{\alpha_{a}} = \alpha_{a}/(4 \pi)$, and $\alpha_a$
are the gauge coupling constants.

\item Scalar mass matrices for sleptons:
%Escalares
\bea
\frac{d (m^{2}_{L})_{ij}}{dt} & = & \delta_{ij} \left( 3
\tilde{\alpha}_2 M_{2}^{2} +  \tilde{\alpha}_1 M_{1}^{2}
\right) -  \frac{1}{16 \pi^{2}} \left[ \frac{1}{2}
(h^{e} {h^{e}}^{\dagger} m^{2}_{L})_{ij} + \frac{1}{2} (m^{2}_{L}
h^{e} {h^{e}}^{\dagger})_{ij} \right. \nonumber \\
& + & \left.
 (h^{e} m^{2}_{e_R} {h^{e}}^{\dagger})_{ij}
+  (m^{2}_{H_{1}} h^{e} {h^{e}}^{\dagger})_{ij}
+ (\hat{A}^{e} {\hat{A}}^{e \dagger})_{ij}
\right]   \; \; , \nonumber\\
\frac{d (m^{2}_{e_R})_{ij}}{dt} & = & \delta_{ij} \left( 4 \tilde{\alpha}_1
M_{1}^{2} \right) -  \frac{1}{16 \pi^{2}} \left[
({h^{e}}^{\dagger} h^{e} m^{2}_{e_R})_{ij} + (m^{2}_{e_R}
{h^{e}}^{\dagger} h^{e})_{ij} \right. \label{rgesc} \\
& + & \left.   2 \left( ({h^{e}}^{\dagger} m^{2}_{L} h^{e})_{ij} +
(m^{2}_{H_{1}} {h^{e}}^{\dagger} h^{e})_{ij} +
({\hat{A}}^{e \dagger} {\hat{A}}^{e})_{ij} \right) \right] \; \; ,
\nonumber
\eea
where $({\hat{A}}^{u,d,e})_{ij} = (A^{u,d,e})_{ij} (h^{u,d,e})_{ij}$.

\item Matricial trilinear terms (also for sleptons):
%Trilineales
\bea
\frac{d (A^{e})_{ij}}{dt} & = & - \delta_{ij} \left( 3 \tilde{\alpha}_2
M_{2} + 3 \tilde{\alpha}_1 M_{1} + \frac{1}{16 \pi^{2}} \left( {\rm Tr}
({\hat{A}}^{e} {h^{e}}^{\dagger})+ 3 {\rm Tr} ({\hat{A}}^{d}{h^{d}}^{\dagger})
\right) \right) \nonumber \\
& - & \frac{1}{16 \pi^{2}}
\left[ 2 ({\hat{A}}^{e} {h^{e}}^{\dagger})_{ij} +
({h^{e}}^{\dagger} {\hat{A}}^{e})_{ij} \right] \;\;. \label{rgestr}
\eea

\item $\mu$ parameter:
%mu
\be
\frac{d \mu^{2}}{dt} = (3 \tilde{\alpha}_2 + \tilde{\alpha}_1) \mu^{2}
- \frac{1}{16 \pi^{2}} {\rm Tr} \left(3 h^{u} {h^{u}}^{\dagger} + 3 h^{d}
{h^{d}}^{\dagger} + h^{e} {h^{e}}^{\dagger} \right) \mu^{2} \;\; .
\label{rgemu}
\ee

\item B parameter:
%Be
\bea
\frac{d B}{dt} & = & - \left( 3 \tilde{\alpha}_2
M_{2} + \tilde{\alpha}_1 M_{1} \right) - \frac{1}{16 \pi^{2}}
{\rm Tr} \left( 3 (h^{u} {\hat{A}}^{u \dagger}) \right.
\nonumber \\
& + & \left. 3 (h^{d} {\hat{A}}^{d \dagger}) +
(h^{e} {\hat{A}}^{e \dagger}) \right) \;\;.
\label{rgeb}
\eea
\end{itemize}

The first thing to notice is that, due to the structure of the equations
for scalar masses (\ref{rgesc}), the ratio $\Delta m^2/m^2$ will in
general be small at low energies (even if it is $O(1)$ at $M_X$),
provided that gaugino masses are bigger\footnote{An evaluation of
the hierarchy which is needed between $M$ and $m$ will be given as a result
of the whole calculation in the following sections.} than scalar masses,
$M_{1/2}^2 \gg m^2$, because of the contribution
of the former in the RGEs of the diagonal parts of the latter, which is not
the case for the off-diagonal entries (note the $\delta_{ij}$ factors in
eqs.~(\ref{rgesc})).

This is the reason why the RGEs have the potential to ``cure'' initial
sizeable values of $\Delta m$. Therefore, the assumption of naturally
large flavour mixing at $M_X$ (see eq.~(\ref{sizedelt})) leads us
necessarily to the $M_{1/2}^2 \gg m^2$ scenario. Moreover, this
sometimes called ``gaugino dominance'' scenario \cite{dine90}
presents other very interesting features, such as that {\em all} the soft
breaking parameters ($M_a$, $m_{ij}^{2}$, $A^{u}_{ij}$, $A^{d}_{ij}$,
$A^{e}_{ij}$ and $B$) are essentially determined {\em at low energies}
by the value of $M_{1/2}$ at $M_X$, independently of their initial
values, as can be again easily seen in eqs.~(\ref{rgegau}, \ref{rgesc},
\ref{rgestr}, \ref{rgeb}). So it is perfectly
sensible to set the $m$s, $A$s and $B$ equal to zero\footnote{Note here
that if $m^2\ll M_{1/2}^2$ at $M_X$, then also $A^2\ll M_{1/2}^2$,
$B^2\ll M_{1/2}^2$, since $A$, $B$ are necessarily $O(m)$ in order
to avoid dangerous charge and color breaking minima. Besides, the value
of $B$ is related to the value of $A$ or $m$ in many supergravity models.
For a review of these issues see \cite{casas95} and references therein.},
and this is what we shall
do for the rest of the analysis. Note that all this does {\em not}
apply to the $\mu$ parameter as it renormalises proportional to itself (see
eq.~(\ref{rgemu})).
So, for a given $M_{1/2}^2 \gg m^2$, the only parameter left from the
initial set (\ref{inPar}) to determine the whole low energy SUSY spectrum is
$\mu$.

However there is a constraint which
we have not considered up to now, namely the requirement of a correct
electroweak breaking. This fixes the value of $\mu$, giving us
the {\em whole} spectrum and other relevant low-energy quantities
(such as $A$ and $\tan\beta$) in terms of a {\em unique}
parameter $M_{1/2}$. From the technical point of view, we have performed the
calculation along the lines already sketched in a previous work
\cite{decar95}, that is,
we have computed the 2--loop RGEs for the gauge couplings and imposed their
unification at $M_X$, taking into account all the SUSY thresholds. Also, to
obtain a correct electroweak breaking, we have included all the spectrum in
the 1-loop effective potential for the Higgs fields\footnote{This
minimization process is
calculated considering only the diagonal mass terms, given the fact that at
the electroweak scale the non-diagonal entries are necessarily small,
precisely due to FCNC constraints like those studied in this paper.},
choosing the appropriate minimization
scale as was already explained in \cite{gambe90,decar93}. In this task
(and also in the rest of
the paper) we work under the assumption that only the Yukawa couplings of the
third generation are the relevant ones, and their values are fixed by the
experimental masses of the corresponding fermions (for the top we take
$m_t=175$ GeV).

The results are summarized in Figs.~2a, 2b.
As a general feature, which is worth mentioning, we have
found that the values of $\tan \beta$ obtained in this framework tend to be
rather large (ranging from 11 to 26 as $M_{1/2}$ increases from
150 GeV to 10 TeV, as can be seen from Fig.~2b); this fact will
be very important in our calculation given the presence of a term
proportional to $\mu\tan \beta$ in one of the contributions to
$BR(\mu\rightarrow e,\gamma)$ (see eqs.~(\ref{amps}) in sect.~3).
We should comment here that
previous analyses have ignored the electroweak breaking constraints,
just fixing $\tan \beta$ to a particular (low) value
\cite{gabbi89,choud95} and making a certain ``average'' in some low-energy
parameters. The present results indicate that this is an important
source of error and uncertainty, which can be avoided by taking into
account the electroweak breaking process.

To summarize, a scenario of gaugino dominance (which arises from the
assumption of naturally large flavour mixing at high energy)
determines the low--energy
spectrum and other relevant low-energy quantities such as
$A$, $\mu$, $\tan\beta$, in terms of a unique parameter, $M_{1/2}$.
This is a very interesting fact that makes
the subsequent analysis rather accurate and model--independent.

\section{Evaluation of $BR( \mu \rightarrow e,  \, \gamma)$}

We follow here a scheme of calculation along the lines of
refs.~\cite{gabbi89,hagel94}.
The effective Lagrangian that describes the
$ \mu \rightarrow e , \gamma$ decay is usually parameterized as
\be
\label{oper}
{\cal L}_{eff} =  \frac{1}{2} \bar{\psi}_{\rm electron} \; \sigma_{\mu \nu}
  (B_L P_L + B_R P_R)  \; \psi_{\rm muon} F^{\mu \nu}\;\;,
\ee
where
%$l$ stands for the muon, $l'$ for the electron and
$P_{R,L} = (1 \pm \gamma_5)/2$. Then, the branching ratio
$BR (\mu \rightarrow e  , \gamma)$ is given by
\be
\label{BR}
BR(\mu \rightarrow e  , \gamma)=\frac{12\pi^2}{G_F^2 m_\mu^2}
\left[|B_L|^2+|B_R|^2\right]\;\;,
\ee
where $G_F$ is the Fermi constant and $m_\mu$ is the muon mass.
The $B_L$, $B_R$ amplitudes arise from one--loop diagrams that
involve a flip of the leptonic flavour triggered by the slepton mixing,
besides the propagation of a neutralino or chargino (see Fig.~1).
Note that the structure of (\ref{oper}) implies that the helicity
of the muon and the electron must be opposite.
Since the electron and muon Yukawa couplings are very suppressed,
only the gauge part of the couplings of the charginos and neutralinos
will play a role in the diagrams. In fact, one
important consequence of the gaugino dominance
framework is that for large enough values of $M_{1/2}$ the
neutralinos (charginos) are almost pure
neutral (charged) gaugino and higgsino.
This result follows from the low-energy inequality
        $M_a^2, \mu^2 \gg M_W^2$,
something that occurs whenever $M_{1/2}\simgt 300$ GeV.
Then, the relevant diagrams, corresponding to
bino ($\tilde{B}$) and wino ($\tilde{W}^{0}$, $\tilde{W}^{-}$) exchange
\footnote{Due to the above-mentioned smallness of the
electron and muon Yukawa couplings,
we have neglected the diagrams involving higgsinos.
Notice also that we have not
included contributions where the helicity flip takes place on
the outgoing electron, because they are suppressed by a $m_e/m_{\mu}$
factor.}, are shown in Fig.~1.
Some comments are in order here. The crosses in the scalar propagators
of the diagrams denote mass insertions. These can be either of the
$ \Delta  m_{ \tilde{e}_L \tilde{\mu}_L }^2,\
 \Delta  m_{ \tilde{e}_R \tilde{\mu}_R }^2 ,\
\Delta  m_{ \tilde{\nu}_e \tilde{\nu}_\mu }^2$ type or of the
$\Delta  m_{ \tilde{\mu_L} \tilde{\mu_R} }^2$ type. The former
occur in all the diagrams and correspond to the off-diagonal
entries in the soft mass matrices of the sleptons. The latter,
occurring only in Diags.~4,6, denotes a change of chirality without
flavour mixing. This arises from
${\cal L}_{\rm SUSY}$ and ${\cal L}_{\rm soft}$
(see eqs.~(\ref{Lsusy},\ref{nodiag})) once $H_1$ and $H_2$
acquire non-vanishing VEVs. It can be checked that
$\Delta  m_{ \tilde{\mu_L} \tilde{\mu_R} }^2=
m_\mu (A^\mu + \mu \tan \beta)$ (note that it is proportional to
the fermion mass, which makes $\Delta  m_{ \tilde{e_L} \tilde{e_R} }^2$
negligible), where from now on we will drop the superindex $\mu$ from
$A^\mu$. It should be noted that the diagrams of Fig.~1 correspond
to an expansion in $\Delta m^2$ (both of
the $ \Delta  m_{ \tilde{e}_L \tilde{\mu}_L }^2,\
 \Delta  m_{ \tilde{e}_R \tilde{\mu}_R }^2 ,\
\Delta  m_{ \tilde{\nu}_e \tilde{\nu}_\mu }^2$ type and of the
$\Delta  m_{ \tilde{\mu_L} \tilde{\mu_R} }^2$ type). This is justified
due to the smallness of $\Delta m^2$ (of either type) compared
to the slepton and gaugino masses. Recall here that even though
$\Delta m^2$ may be $O(m^2)$ at $M_X$, it must necessarily be
$\ll m^2$ at low energies in order to be consistent with the experiment
(this will become apparent in the next section). On the other hand
the crosses in the fermionic propagators denote a change in helicity,
which gives a factor proportional to the fermion mass.
Let us remark here that this emerges from the pure calculation
and does not amount to any mass expansion.

Splitting the induced $B_{L,R}$ values as
\bea
\label{descBs}
B_R & = &  B^{\tilde{W}^-}_R + B^{\tilde{W}^0}_R +  B^{\tilde{B}}_R \;,
\nonumber\\
B_L & = & B^{\tilde{B}}_L\;,
\eea
%\phantom{B^{W^-}_L + B^{W^0}_L         +} B^{\tilde{B}}_L \;\;,
%
the different diagrams contribute in the following way:
\bea
\label{BsDs}
B^{\tilde{W}^-}_R &=& {\cal D}_1 \;,\hspace{2.3cm}
B^{\tilde{W}^0}_R = {\cal D}_2 \;,
\nonumber\\
B^{\tilde{B}  }_R &=& {\cal D}_3 + {\cal D}_4 \;,\hspace{1.3cm}
B^{\tilde{B}  }_L = {\cal D}_5 + {\cal D}_6\;\;,
\eea
where ${\cal D}_1-{\cal D}_6$ correspond to diagrams 1--6 of Fig.~1.
We have evaluated {\em all} of them.
The expressions given in the previous literature
are either incomplete or not directly applicable to our case.
Some of them correspond to the limit in which the photino
exchange gives the main contribution (which is not valid in
this case). Other estimates do not take into account the
left-right mixing between charged sleptons, that will be
essential, or ignore some of the diagrams.
Also, it is normally assumed that left and right sleptons
are degenerated in mass\footnote{These incomplete expressions in
the literature should however
coincide with a piece of our calculation in the appropriate limits.
In this sense, we obtain a complete agreement with ref.~\cite{gabbi89}
while, regarding ref.~\cite{hagel94}, we find an additional factor 2
in ${\cal D}_4$ and a different relative sign between ${\cal D}_1$
and ${\cal D}_2$ (see eqs.~(\ref{Ds}) below).}. However, it is clear
that a mass splitting will usually arise due to the different
dependence of the RGEs, eq.~(\ref{rgesc}), on the gaugino masses.
In the gaugino dominance scenario this effect tends to be very large,
see Fig.~2a.

The complete expressions are:
\bea
\label{Ds}
  {\cal D}_1 & = &
               \frac {e^3} {16 \pi^2}
               \frac {m_{\mu}} { \msn^2}
               \frac {\Delta  m_{ \tilde{\nu}_e \tilde{\nu}_\mu }^2  }
                     {\msn^2}
               \frac {G ( \frac { M^2_{\tilde{W}^-} }  { \msn^2} )}
                     {\sin^2 \theta_W} \nonumber \\
  {\cal D}_2 & = &
             - \frac {e^3} {16 \pi^2}
               \frac {m_{\mu}} {\msl^2}
               \frac {\Delta  m_{ \tilde{e}_L \tilde{\mu}_L }^2  }
                     {\msl^2}
               \frac {F ( \frac { M^2_{\tilde{W}^0} }
                                     { \msl^2            }
                        )                             }
                     {2 \sin^2 \theta_W} \nonumber \\
  {\cal D}_3 & = &
             -  \frac {e^3}  {16 \pi^2}
                \frac {m_\mu}  {\msl^2}
                \frac {\Delta  m_{ \tilde{e}_L \tilde{\mu}_L }^2  }
                      {\msl^2}
                \frac {F  ( \frac { M^2_{\tilde{B}} }
                                      { \msl^2          }  ) }
                      {2 \cos^2 \theta_W} \label{amps} \nonumber\\
  {\cal D}_4 & = &
                \frac {e^3}  {16 \pi^2}
                \frac {M_{\tilde{B}} m_\mu (A + \mu \tan \beta)}
                      {(\msl^2 - \msr^2)}
   \left[       \frac {\Delta  m_{ \tilde{e}_L \tilde{\mu}_L }^2  }
                      {(\msl^2 - \msr^2)}
         \left(
                \frac { F ( \frac { M^2_{\tilde{B}} }
                                       { \msl^2          } ) }
                      {\msl^2} -
                \frac { F ( \frac { M^2_{\tilde{B}} }
                                       { \msr^2          } ) }
                      {\msr^2}
         \right) \right. \nonumber \\
             & + & \left. \frac {\Delta  m_{ \tilde{e}_L \tilde{\mu}_L }^2  }
                      {\msl^2}
                \frac {L ( \frac { M^2_{\tilde{B}} }
                                      { \msl^2          }  ) }
                      {\msl^2}
   \right]
                \frac {1}{ \cos^2 \theta_W} \nonumber \\
  {\cal D}_5 & = &
             -  \frac {e^3}  {16 \pi^2}
                \frac {m_\mu}  {\msr^2}
                \frac {\Delta  m_{ \tilde{e}_R \tilde{\mu}_R }^2  }
                      {\msr^2}
                \frac {2 F ( \frac { M^2_{\tilde{B}} }
                                      { \msr^2          }  ) }
                      {\cos^2 \theta_W} \nonumber \\
   {\cal D}_6 & = &
                \frac {e^3}  {16 \pi^2}
                \frac {M_{\tilde{B}} m_\mu (A + \mu \tan \beta)}
                      {(\msr^2 - \msl^2)}
   \left[       \frac {\Delta  m_{ \tilde{e}_R \tilde{\mu}_R }^2  }
                      {(\msr^2 - \msl^2)}
         \left(
                \frac { F ( \frac { M^2_{\tilde{B}} }
                                       { \msr^2          } ) }
                      {\msr^2} -
                \frac { F ( \frac { M^2_{\tilde{B}} }
                                       { \msl^2          } ) }
                      {\msl^2}
         \right) \right. \nonumber \\
             & + & \left. \frac {\Delta  m_{ \tilde{e}_R \tilde{\mu}_R }^2  }
                      {\msr^2}
                \frac {L ( \frac { M^2_{\tilde{B}} }
                                      { \msr^2          }  ) }
                      {\msr^2}
   \right]
                \frac {1}{ \cos^2 \theta_W}
\eea
where
\bea
\label{emes}
m_{ \tilde{\nu}_e }^2   =  m_{ \tilde{\nu}_\mu }^2   \equiv  \msn^2\, ,  \;\;
m_{ \tilde{e}_L   }^2   =  m_{ \tilde{\mu}_L   }^2  \equiv \msl^2\, ,  \;\;
m_{ \tilde{e}_R   }^2   =  m_{ \tilde{\mu}_R   }^2  \equiv  \msr^2
\eea
and
\bea
F(x) & = & \frac{1}{12} (1-x)^{-5}
       \left[ 17x^3 - 9x^2 - 9x + 1 - 6x^2(x+3)\ln x \right]
\nonumber\\
G(x) & = &\frac{1}{6}  (1-x)^{-5}
       \left[ - x^3 - 9x^2 + 9x     + 6x  (x+1)\ln x \right]
\nonumber\\
L(x) & = & \frac{1}{2}  (1-x)^{-4}
       \left[       - 5x^2 + 4x + 1 + 2x  (x+2)\ln x \right]
\label{Fs}
\eea
Note that the equalities of eq.~(\ref{emes}) do not come
from any universality assumption, but from the mere gaugino dominance
in the corresponding RGEs.

Although in principle all the diagrams can have a similar magnitude
(e.g. if we assume
$\Delta  m_{ \tilde{\nu}_e \tilde{\nu}_\mu }^2 \sim
 \Delta  m_{ \tilde{e}_L \tilde{\mu}_L }^2 \sim
 \Delta  m_{ \tilde{e}_R \tilde{\mu}_R }^2 $), in practice diagrams 4
and 6 are the dominant ones.
This comes from the coefficient of proportionality $(A+\mu\tan\beta)$,
which turns out to be very important in the gaugino dominance framework
due to the large $\tan\beta$ value (see sect.~2).

Finally, let us remark that the previous results on $BR(\mu\rightarrow
e,\gamma)$ are in agreement with the recent paper of ref.~\cite{dimop95},
where charginos and neutralinos are allowed to be non-pure gaugino
or higgsino states.

\section{Results}

The expressions obtained in the previous section for
$BR(\mu \rightarrow e , \gamma)$ depend on two different
sets of parameters. First, the different masses involved in the game
($\msn^2, \; \msl^2, \; \msr^2,$ $\; M_{\tilde{B}}, \; M_{\tilde{W}}$)
and certain relevant  low-energy quantities $(A, \mu, \tan \beta)$.
Second, the three independent flavour-mixing mass entries:
$\Delta  m_{ \tilde{\nu}_e \tilde{\nu}_\mu }^2 , \;
 \Delta  m_{ \tilde{e}_L \tilde{\mu}_L }^2     , \;
 \Delta  m_{ \tilde{e}_R \tilde{\mu}_R }^2 $.
As explained in sect.~2, once we are working in the framework of gaugino
dominance, $M_{1/2}^2 \gg m^2 $, the first set is completely determined in
terms of the initial gaugino mass, $M_{1/2}$. Recall that we were led to
this framework by the mere assumption of naturally large flavour mixing at
$M_X$ (see eq.~(\ref{sizedelt})). The three flavour-mixing mass parameters,
however, remain independent (and we will consider them in that way in
the following), although it is logical to suppose that they are of the
same order.

The constraints on the MSSM from $BR(\mu \rightarrow e , \gamma)$
arise by comparing the above calculations (eqs.~(\ref{BR}--\ref{Fs}))
with the present
experimental bound, eq.~(\ref{megexp}). We have illustrated this in
Fig.~3, where an overall mass-mixing parameter
$\Delta  m_{ \tilde{\nu}_e \tilde{\nu}_\mu }^2  \, = \,
 \Delta  m_{ \tilde{e}_L \tilde{\mu}_L }^2      \, = \,
 \Delta  m_{ \tilde{e}_R \tilde{\mu}_R }^2      \,\equiv \, \Delta  m^2 $
has been taken for simplicity. Then we have plotted
$BR(\mu\rightarrow e,\gamma)$ vs.
$M_{1/2}$ for different values of $\Delta m$.
{}From this figure we can derive the maximum allowed value of $\Delta m$
(or, equivalently, the minimum allowed value of
$M_{1/2} / \Delta m$) for each value of
$M_{1/2}$. This is represented in Fig.~4 for four different cases:
{\em a)} $\Delta  m_{ \tilde{\nu}_e \tilde{\nu}_\mu }^2  \, = \,
 \Delta  m_{ \tilde{e}_L \tilde{\mu}_L }^2        \, = \,
 \Delta  m_{ \tilde{e}_R \tilde{\mu}_R }^2        \, \equiv \, \Delta  m^2 $;
{\em b)} only
$ \Delta  m_{ \tilde{e}_R \tilde{\mu}_R }^2       \, \ne \, 0$;
{\em c)} only
$ \Delta  m_{ \tilde{e}_L \tilde{\mu}_L }^2       \, \ne \, 0$
and
{\em d)} only
$ \Delta  m_{ \tilde{\nu}_e \tilde{\nu}_{\mu} }^2 \, \ne \, 0$,
which gives a complete picture of the results.
Notice that the {\em (d)} case is the less restrictive one.
This is because, as explained in sect.~3, given the large value of
$\tan\beta$ the dominant diagrams are nos. 4 and 6 of Fig.~1,
which do not involve
$ \Delta  m_{ \tilde{\nu}_e \tilde{\nu}_{\mu} }^2  $, see eq.~(\ref{Ds}).
On the other hand, the strongest constraint comes
from $ \Delta  m_{ \tilde{e}_R \tilde{\mu}_R }^2 $.
This is because the mass of the right sleptons, $\msr^2$, is smaller
than that of the left ones, $\msl^2$, as a consequence of their
different RG running (note from eq.~(\ref{rgesc}) the different dependence
of $\msl^2$, $\msr^2$ on the gaugino masses).

The constraints are in general extremely strong. For
case {\em (a)}, which is the most representative one, the corresponding
curve can be approximately fitted by the simple
constraint\footnote{A more accurate form for the constraint
is $\frac{M_{1/2}^2}{\Delta m}\geq (9\ {\rm TeV})\sqrt{\tan \beta}$,
where $\tan\beta$ can be well fitted by $\tan\beta=16\left[
M_{1/2} ({\rm TeV})\right]^{1/5}$, see Fig.~2b. The dependence
of $\frac{M_{1/2}^2}{\Delta m}$ on $\sqrt{\tan \beta}$ can be
understood from the dependence of the dominant diagrams (nos. 4 and 6
of Fig.~1) on $\tan\beta$, see eqs.~(\ref{Ds}).}
\be
\label{fit}
\frac{M_{1/2}^2}{\Delta m}\simgt 34\ {\rm TeV}
\ee
(similar equations can be written for the curves associated to
the other {\em (b)}, {\em (c)}, {\em (d)} cases). Under the
assumption of eq.~(\ref{sizedelt}), i.e. $\Delta m = O(m)$, the results
of Fig.~4 or eq.~(\ref{fit}) imply that, indeed, a very large
hierarchy between the scalar and gaugino masses is needed in order
to reconcile the theoretical and experimental results. This
gives full justification to our assumption of a gaugino dominance
framework once eq.~(\ref{sizedelt}) has been conjectured. For
example, for $M_{1/2}\sim 500$ GeV the assumption $\Delta m\sim m$
demands $M_{1/2}/\Delta m> 65$. For smaller values of
$M_{1/2}$ the required hierarchy is in fact more severe, whereas
too large values of $M_{1/2}$ start to conflict an electroweak
breaking process with no fine--tuning \cite{decar93,barbi88}.
Actually, it is hard to think of a scenario where such a dramatical
hierarchy can naturally arise. Consequently, we can conclude at this
point that a naturally large flavour mixing, as that conjectured in
eq.~(\ref{sizedelt}), can hardly be reconciled with the experiment in
a natural way.

On the other hand, the results from Fig.~4 and eq.~(\ref{fit}) are so
extreme that we can relax the assumption $\Delta m \sim m$, allowing
for larger values of $m$ relative to $\Delta m$ without losing the
validity of the calculation. The latter is based on the use of the
gaugino dominance framework, which is reasonably accurate for
$M_{1/2}^2 > O(10)\ m^2$.
Then, from (\ref{fit}), it follows that all our results, and
eq.~(\ref{fit}) itself, are valid whenever we start with a $\Delta m$
at $M_X$ such that
\be
\label{fit2}
\frac{\Delta m^2}{m^2}\simgt \frac{m^2}{10\ {\rm TeV^2}}\;,
\ee
as was foretold in eq.~(\ref{sizedelt2}). This result extends enormously
the scope of application of our calculation and hence its interest.
In this way we see that rather small values of $\Delta m^2/m^2$
at $M_X$ require a gaugino dominance scenario to be cured, thus leading to
a complete determination of all the relevant low-energy parameters
in terms of $M_{1/2}$ (see sect.~2 and Fig.~2).

It is interesting to compare these results with those obtained in
the initial calculation of ref.~\cite{choud95}. There, besides assuming
$ \Delta  m_{ \tilde{e}_L \tilde{\mu}_L }^2 \neq 0$,
$ \Delta  m_{ \tilde{e}_R \tilde{\mu}_R }^2 =
\Delta  m_{ \tilde{\nu}_e \tilde{\nu}_\mu }^2=0$ (i.e. our case
{\em (c)}), several additional assumptions and averages were adopted
(some of them already commented in sect.~2). The confrontation of the
two calculations is made in Fig.~5, where we have represented
$\left(M_{1/2}/\Delta m \right)_{\rm min}$ vs. $m_{\rm av}
\equiv\sqrt{\frac{1}{2}(\msl^2+\msr^2)}$ to facilitate
the comparison with the plots of ref.~\cite{choud95}. Obviously,
there is a big difference between their results (that we have been able
to reproduce) and ours. This reflects the very large uncertainties
in the calculation of ref.~\cite{choud95} (acknowledged by its authors),
which are dramatically suppressed once a correct electroweak breaking
process is imposed.

\section{Summary and concluding remarks}

In the absence of any additional assumption it is natural to conjecture
that sizeable flavour-mixing mass entries, $\Delta m^2$, may appear in the
mass matrices of the scalars of the MSSM, i.e. $\Delta m^2\sim O(m^2)$.
This flavour violation can still be reconciled with the experiment
if the gaugino mass, $M_{1/2}$, is large enough to yield (through the
renormalization group running) a sufficiently small $\Delta m^2 / m^2$
at low energy. We have analyzed in detail this possibility, focussing
our attention on the leptonic sector, particularly on the
$\mu\rightarrow e,\gamma$ decay, which is by far the FCNC process with
higher potential to restrict the value of the off-diagonal terms,
$\Delta m^2$. The results are the following:

\begin{enumerate}

\item The $\Delta m^2\sim O(m^2)$ conjecture automatically leads
to a {\em gaugino dominance} framework (i.e. $M_{1/2}^2\gg m^2$), where,
apart from $\Delta m^2$ itself, {\em all} the
relevant low-energy quantities (mass spectrum, $A$,
$\mu$, $\tan\beta$) are determined in terms of a unique parameter,
$M_{1/2}$ (see Fig.~2).
This makes the subsequent analysis and results remarkably
model--independent.

\item The resulting constraints in the MSSM, obtained by comparing
the calculated $BR(\mu\rightarrow e,\gamma)$ with the experimental
bound, are very strong (see Figs.~3, 4). More precisely, assuming
$\Delta  m_{ \tilde{\nu}_e \tilde{\nu}_\mu }^2  \, = \,
 \Delta  m_{ \tilde{e}_L \tilde{\mu}_L }^2  \, = \,
 \Delta  m_{ \tilde{e}_R \tilde{\mu}_R }^2 \, \equiv \, \Delta  m^2 $,
we arrive at the approximate constraint
\be
\label{fit2p}
\frac{M_{1/2}^2}{\Delta m}\simgt 34\ {\rm TeV}
\ee
(and similar equations for other cases). This makes, in
our opinion, the natural flavour
mixing conjecture $\Delta m^2\sim O(m^2)$ extremely hard to be
reconciled with the experiment in a natural way.
Hence, $\Delta m/m$ should be
small already at the unification scale.

\item The required values of  $M_{1/2}/\Delta m$ to
be within the experimental limits are so large that the
gaugino dominance assumption, which is an essential ingredient
of our analysis,
remains valid for values of $m^2$ much larger than $\Delta m^2$,
namely for
$\frac{\Delta m^2}{m^2}\simgt \frac{m^2}{10\ {\rm TeV^2}}$,
thus extending enormously
the scope of interest and application of our results.

\end{enumerate}

\noindent
To perform the previous calculation we have completed the earlier
evaluations~\cite{gabbi89,hagel94} of $BR(\mu\rightarrow e,\gamma)$
(see diagrams of Fig.~1). The results, summarized in
eqs.~(\ref{BR}--\ref{Fs}), are in concordance with the recent results
of ref.~\cite{dimop95}.

We would like to stress the high degree of model--independence of our
analysis and results. In fact, in the entire calculation we have only
made, for the sake of simplicity, the assumption of universality of
gaugino masses at the unification scale. The relaxation of this
assumption does not imply any essential conceptual change in the
analysis, leading to straightforward modifications in the results of
the paper, without affecting the main conclusions above.

Finally, let us comment that the need of starting with small
$\Delta m/m$ can be satisfied in some theoretically well-founded
scenarios, which become favoured from this point of view. In
particular, besides the proposed mechanisms of
refs.~\cite{nir93,dine93,dimop95p}, we would like to stress that many
string constructions can be consistent with that requirement. More
precisely, in orbifold compactifications schemes \cite{dixon86}, which
are known to give an interesting phenomenology \cite{casas88},
the soft SUSY breaking scalar masses always consist of a universal
piece plus a contribution proportional to the so-called modular
weight ($n$) associated with the field under consideration
\cite{ibane92,decar93p}.
Thus, scalar fields belonging to the same twisted sector of the
theory (or to the untwisted one) acquire degenerate soft masses,
something that can perfectly occur in many realistic scenarios.
A similar thing happens in the so-called dilaton-dominated limit
\cite{kaplu93,brign94}.
Other scenarios, however, can produce a larger non-universality
of the scalar masses, with potentially dangerous contributions to FCNC
processes \cite{brax94}. In any case, these non-universality effects
are to produce non-vanishing off-diagonal terms in the scalar
mass matrices once the usual rotation of fields to get diagonal
fermionic mass matrices is carried out. The
phenomenological viability of these physically relevant scenarios
undoubtedly deserves further investigation. Work along these
lines is currently in progress \cite{decar96}.

\section*{Acknowledgements}

We thank S. Pokorski for his help to clarify some aspects of
ref.~\cite{choud95}.

\newpage

%%%%%%%%%%%%%%%%%%%%%%%%FIGURA%%%%%%%%%%%%%%%%%%%%%%%%
\begin{figure}
\centerline{
\psfig{figure=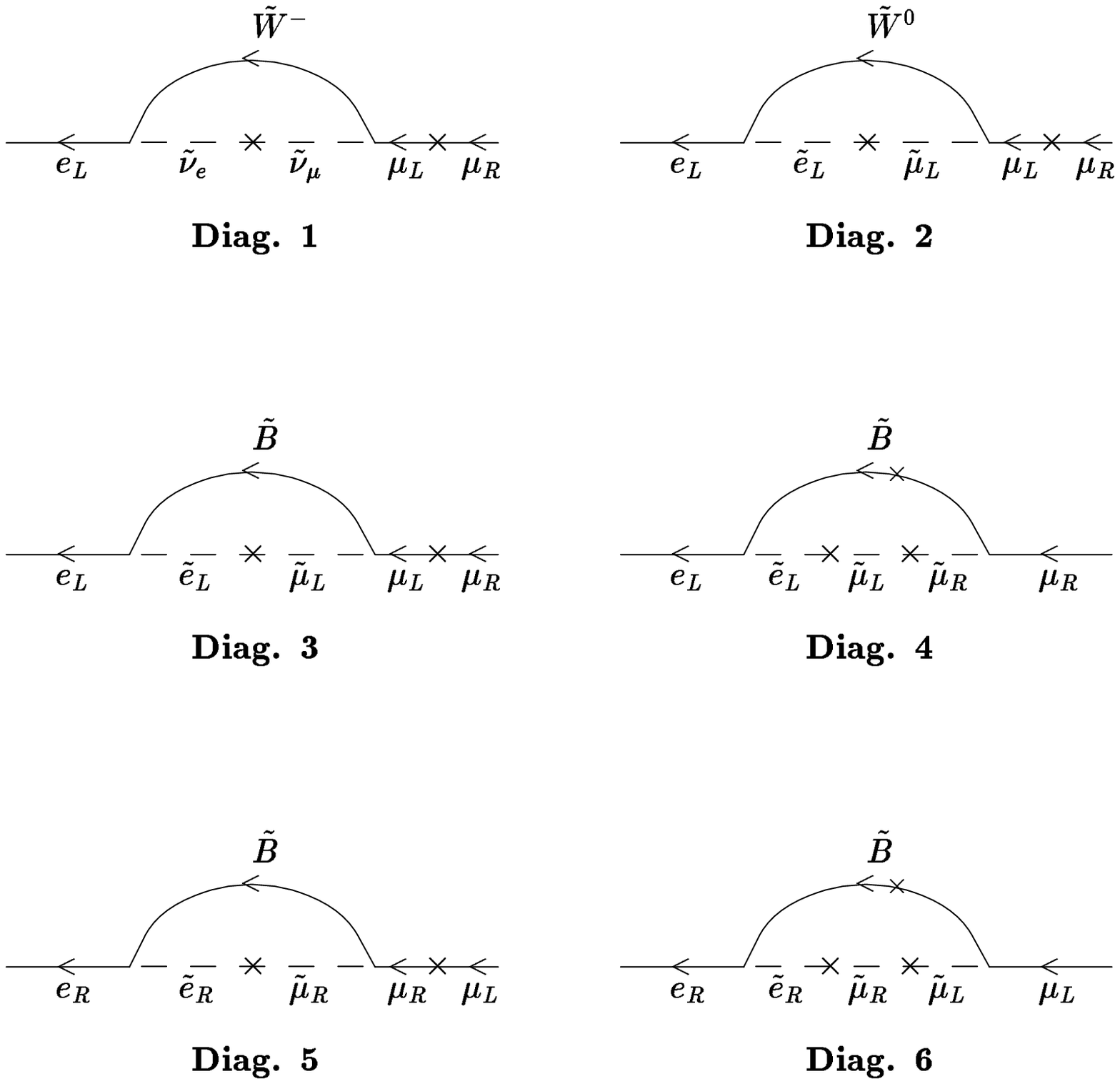,height=12cm,bbllx=2.cm,bblly=7.cm,bburx=21.cm,bbury=21cm}}
\caption{\tenrm\baselineskip=10pt Supersymmetric contributions to
$\mu\rightarrow e,\gamma$.
The crosses indicate mass insertions (for scalars) and helicity flips
(for fermions). The outgoing photon (not represented in the graphs)
is assumed to be attached to the different diagrams in all possible ways.}
\end{figure}
%%%%%%%%%%%%%%%%%%%%%%%%%%%%%%%%%%%%%%%%%%%%%%%%%%%%%%
\begin{figure}
\centerline{\vbox{
\psfig{figure=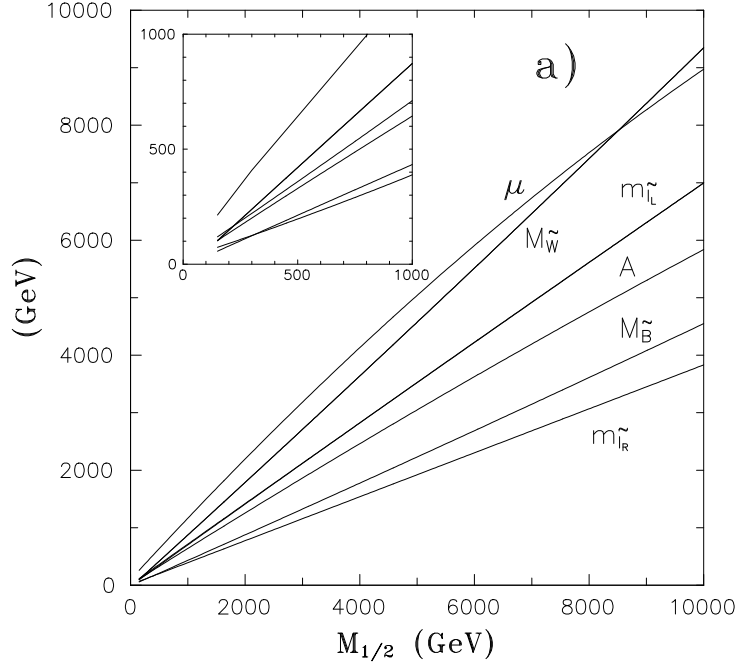,height=10cm,bbllx=4.cm,bblly=2.cm,bburx=14cm,bbury=14cm}
\psfig{figure=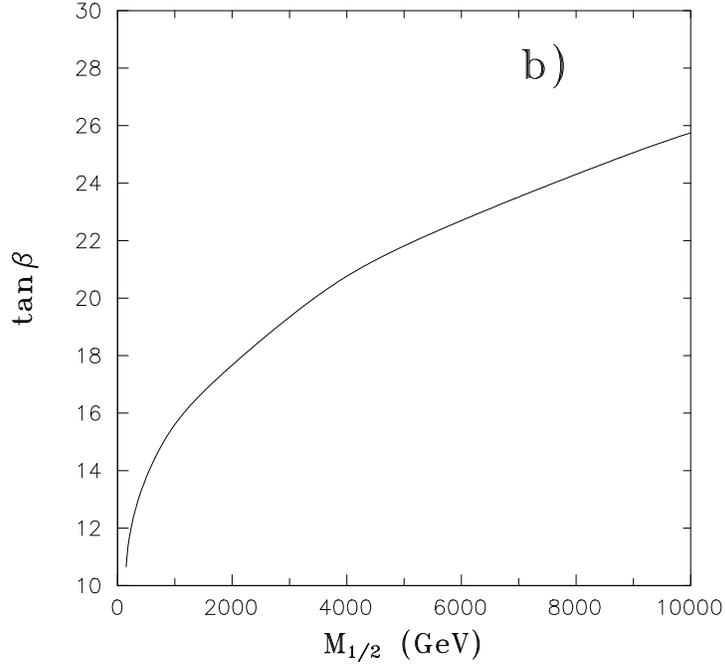,height=10cm,bbllx=4.cm,bblly=2.cm,bburx=14cm,bbury=14cm}
}}
\caption{\tenrm\baselineskip=10pt a) Relevant mass spectrum ($M_{\tilde{W}}$,
$M_{\tilde{B}}$, $\msl$, $\msr$, $\msn$) and significant low-energy
quantities ($A$, $\mu$) in the gaugino dominance framework
($M_{1/2}^2\gg m^2$) as functions of $M_{1/2}$ (the only free
parameter). The slight non-degeneracies between
($M_{\tilde{W}^{0}}$, $M_{\tilde{W}^{\pm}}$) and between
($m_{\tilde{e}_L}$, $m_{\tilde{\nu}_L}$) have not been represented.
The curves are cut at $M_{1/2}=$150 GeV, where the
spectrum starts to conflict with the experimental bounds.
b) Plot of $\tan\beta$ vs. $M_{1/2}$ in the gaugino dominance framework.}
\end{figure}
%%%%%%%%%%%%%%%%%%%%%%%%%%%%%%%%%%%%%%%%%%%%%%%%%%%%%%
\begin{figure}
\centerline{\vbox{
\psfig{figure=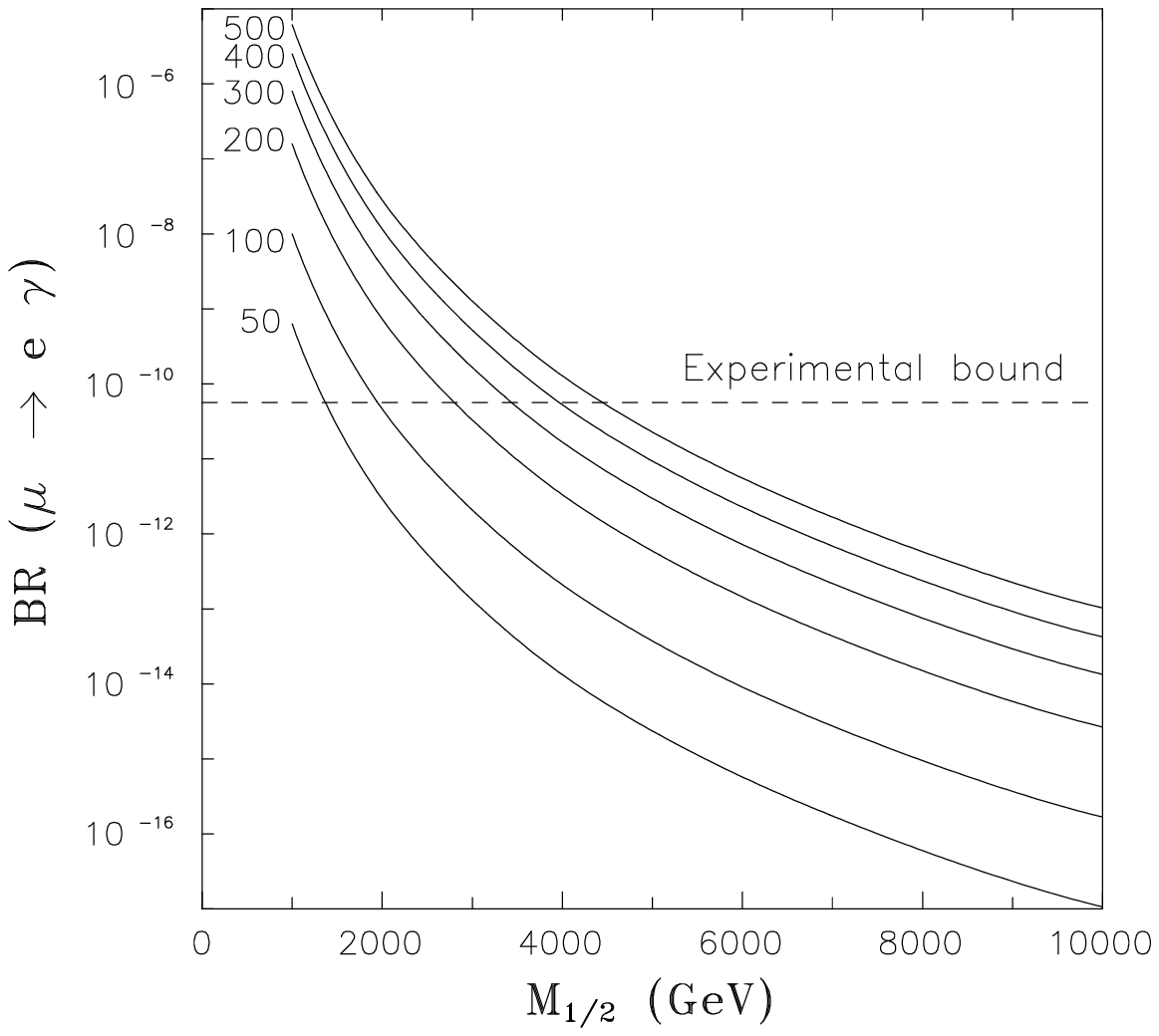,height=9.5cm,bbllx=0.5cm,bblly=3.cm,bburx=13.5cm,bbury=14.5cm}
\caption{\tenrm\baselineskip=10pt Plot of $BR(\mu\rightarrow
e,\gamma)$ vs. $M_{1/2}$,
taking for simplicity $\Delta  m_{ \tilde{\nu}_e \tilde{\nu}_\mu }^2 =
\Delta  m_{ \tilde{e}_L \tilde{\mu}_L }^2  = \Delta  m_{ \tilde{e}_R
\tilde{\mu}_R }^2 \equiv  \Delta  m^2 $.
The different curves correspond to $\Delta m = $ 50, 100, 200, 300, 400,
500 GeV respectively.}
\psfig{figure=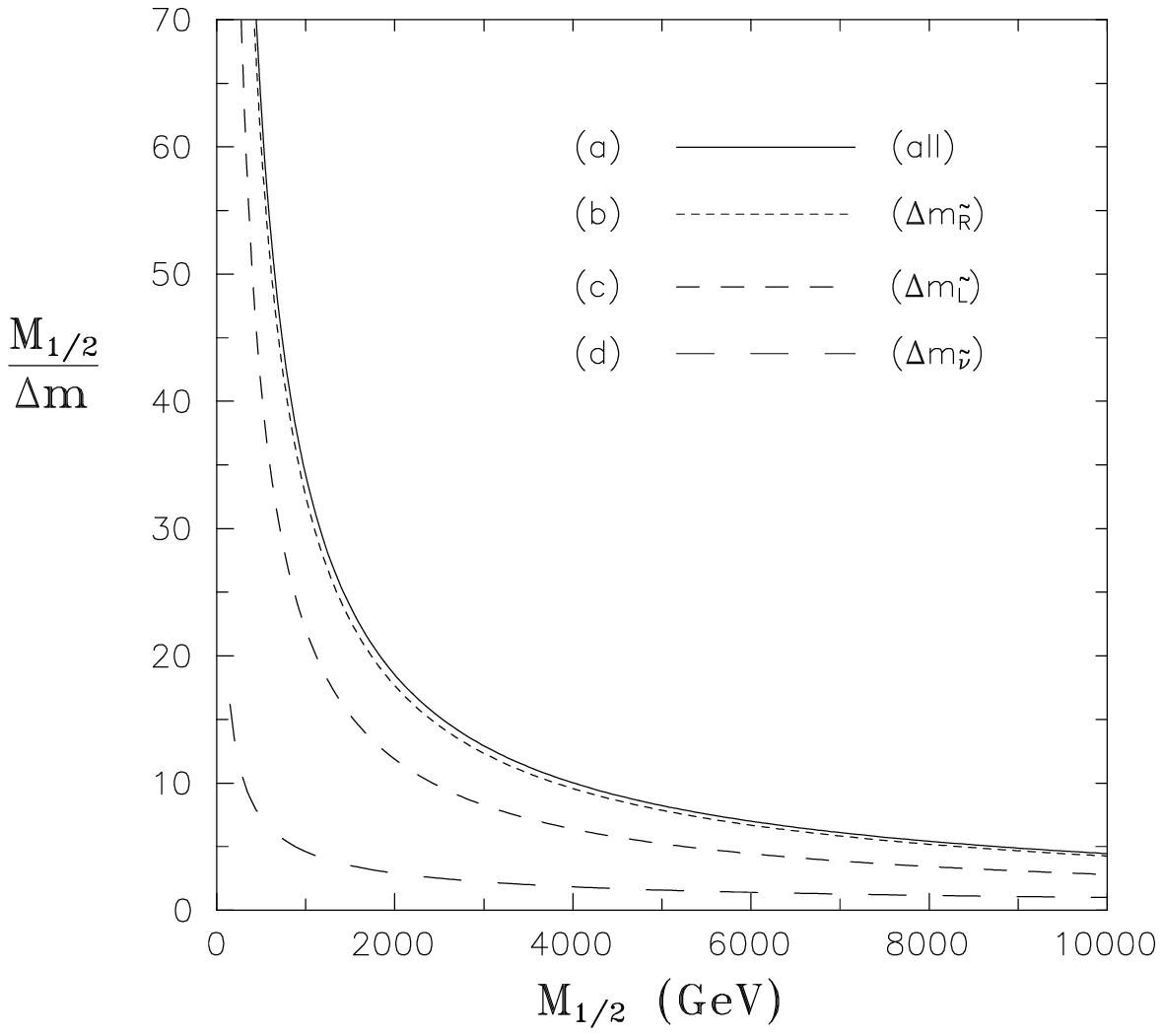,height=9.5cm,bbllx=0.5cm,bblly=3.cm,bburx=13.5cm,bbury=14.5cm}
\caption{\tenrm\baselineskip=10pt
Plot of the {\em minimum} allowed
value of $M_{1/2}/\Delta m$ vs. $M_{1/2}$ in four different cases:
{\em a)} $\Delta  m_{ \tilde{\nu}_e \tilde{\nu}_\mu }^2  \, = \,
 \Delta  m_{ \tilde{e}_L \tilde{\mu}_L }^2        \, = \,
 \Delta  m_{ \tilde{e}_R \tilde{\mu}_R }^2        \, \equiv \, \Delta  m^2 $;
{\em b)} only
$ \Delta  m_{ \tilde{e}_R \tilde{\mu}_R }^2       \, \ne \, 0$;
{\em c)} only
$ \Delta  m_{ \tilde{e}_L \tilde{\mu}_L }^2       \, \ne \, 0$
and
{\em d)} only
$ \Delta  m_{ \tilde{\nu}_e \tilde{\nu}_\mu }^2 \, \ne \, 0$.}}}
\end{figure}
%%%%%%%%%%%%%%%%%%%%%%%%%%%%%%%%%%%%%%%%%%%%%%%%%%%%%%
\begin{figure}
\centerline{
\psfig{figure=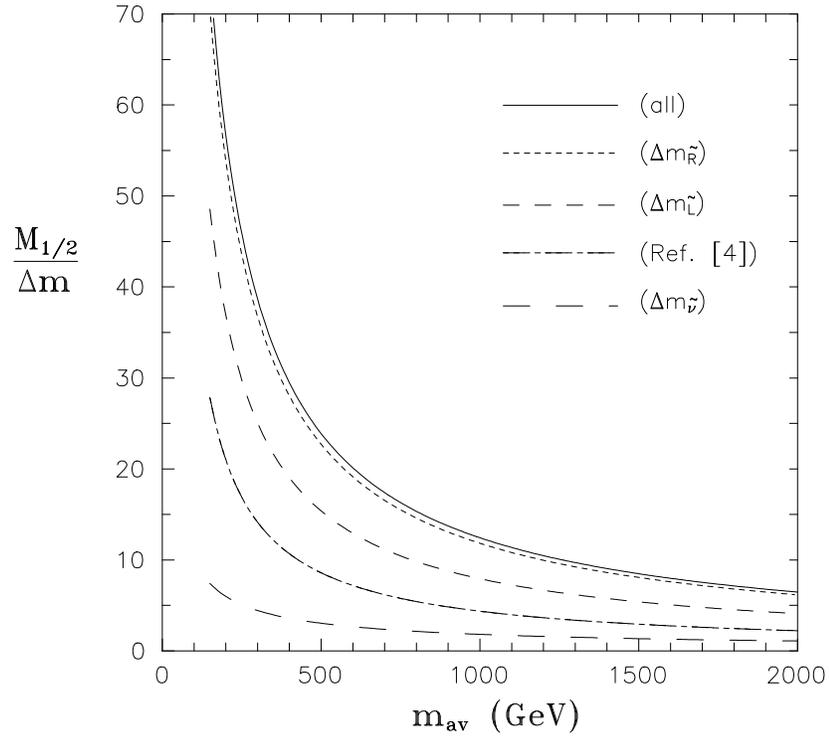,height=12cm,bbllx=3.cm,bblly=2.cm,bburx=15cm,bbury=15cm}
}
\caption{\tenrm\baselineskip=10pt The same as in Fig.~4, but
representing
$m_{\rm av} \equiv \left(\frac{1}{2}(\msl^2+\msr^2) \right)^{1/2}$
in the horizontal axis. The
dashed--dotted line corresponds to the calculation of ref.~[4].}
\end{figure}
%%%%%%%%%%%%%%%%%%%%%%%%%%%%%%%%%%%%%%%%%%%%%%%%%%%%%%

\end{document}